\documentclass{aa}
\usepackage{graphicx}
\usepackage{txfonts}
\usepackage{natbib}
\usepackage{epsfig}

\begin{document}


\title{The outburst of the eruptive young star OO\,Serpentis
    between 1995 and 2006\thanks{Based on observations with ISO, an
    ESA project with instruments funded by ESA member states
    (especially the PI countries France, Germany, the Netherlands and
    the United Kingdom) with participation of ISAS and NASA.}}
%
%
\author{\'A.~K\'osp\'al\inst{1}
        \and
        P.~\'Abrah\'am\inst{1}
        \and
        T.~Prusti\inst{2}
        \and
        J.~Acosta-Pulido\inst{3}
        \and
        S.~Hony\inst{4}
        \and
        A.~Mo\'or\inst{1}
        \and
        R.~Siebenmorgen\inst{5}}
\offprints{kospal@konkoly.hu}
\institute{Konkoly Observatory of the Hungarian Academy of Sciences, P.O.Box
           67, H-1525 Budapest, Hungary 
\and
           ESTEC/SCI-SAF, Postbus 299, 2200 AG Noordwijk, The
           Netherlands
\and
           Instituto de Astrof\'\i{}sica de Canarias, 38205 La Laguna,
           Tenerife, Canary Islands, Spain
\and       Instituut voor Sterrenkunde, K.U. Leuven, Celestijnenlaan 200B, B-3001
           Leuven, Belgium 
\and
           European Southern Observatory, Karl-Schwarzschild-Strasse 2,
           85748 Garching, Germany}
\date{Received date; accepted date}
\titlerunning{The 1995--2004 outburst of OO\,Serpentis}

\abstract{}
{OO\,Serpentis is a deeply embedded pre-main sequence star in the
Serpens NW star-forming region. The star went into outburst in 1995
and gradually faded afterwards. In many respects its eruption
resembled the well-known FU Orionis-type (FUor) or EX Lupi-type (EXor)
outbursts. Since very few such events have ever been documented at
infrared wavelengths, our aim is to study the temporal evolution of
OO\,Ser in the infrared.}
{OO\,Ser was monitored with the \emph{Infrared Space Observatory} in
the $3.6\,{-}\,100\,\mu$m wavelength range, starting 4 months after
peak brightness and covering a period of 20 months. Eight years later,
in 2004-2006 we again observed OO\,Ser at $2.2$ and $12\,\mu$m from
the ground and complemented this dataset with archival Spitzer
obsevations also from 2004. We analysed these data with special
attention to source confusion and constructed light curves at 10
different wavelengths as well as spectral energy distributions.}
{The outburst caused brightening in the whole infrared regime.
According to the infrared light curves, OO\,Ser started a
wavelength-independent fading after the peak brightness. Later the
flux decay became slower but stayed practically
wavelength-independent. The fading is still ongoing, and current
fading rates indicate that OO\,Ser will not return to quiescent state
before 2011. The outburst timescale of OO\,Ser seems to be shorter
than that of FUors, but longer than that of EXors.}
{The outburst timescale and the moderate luminosity suggest that
OO\,Ser is different from both FUors and EXors, and shows some
similarities to the recently erupted young star V1647\,Ori. Based on
its SED and bolometric temperature, OO\,Ser seems to be an early class
I object, with an age of ${<}\,10^5\,$yr. As proposed by
outburst models, the object is probably surrounded by an accretion
disc and a dense envelope. This picture is also supported by the
wavelength-independence of the fading. Due to the shorter outburst
timescales, models developed for FUors can only work for OO\,Ser if
the viscosity parameter in the circumstellar disc, $\alpha$, is set to
an order of magnitude higher value than usual for FUors.}


\maketitle


\section{Introduction}
\label{sec:intro}

OO\,Serpentis ($\alpha_{2000}\,{=}\,18^{\rm h} 29^{\rm m} 49\,\fs1$,
$\delta_{2000}\,{=}\,{+}01\degr{} 16\arcmin{} 20\arcsec{}$) is a
deeply embedded pre-main sequence star in the Serpens NW star-forming
region at a distance of 311 pc. In 1995 \object{OO\,Ser} underwent a
large increase in K-band flux in less than 1 year, reaching its
maximum brightness in 1995 October \citep{hodapp96}. The object was
not visible, even at peak brightness, in the J-band or at shorter
wavelengths, therefore it is also known as Serpens Deeply Embedded
Outburst Star (DEOS). In the H and K bands the object was observable
but the emission is dominated by scattered light. In the K-band
\citet{hodapp99} monitored the outburst until 1998 October, and found
that after the peak OO\,Ser gradually faded at a rate faster than the
typical fading rate of FU Orionis-type objects (FUors), but slower
than that of EX Lupi-type stars (EXors). Its K-band spectrum (a
steeply rising, smooth continuum) also differed from both FUor and
EXor spectra, which usually exhibit absorption or emission features.

Eruptions of pre-main sequence stars are rare events, thus a new
outburst is always noteworthy. The eruption mechanism is thought to be
caused by enhanced accretion from the circumstellar disc onto the star
\citep[e.g.~][]{hk96}. The close link between the eruption phenomenon
and the circumstellar material makes it crucial to document the
outburst also at infrared wavelengths, where the circumstellar dust
radiates. However, such observing programmes are constrained by the
availability of active infrared satellite missions. The physical
analysis of the phenomenon is limited due to the lack of preoutburst
data. The eruption of OO\,Ser in 1995 provided a unique opportunity to
collect such a dataset and carry out a multiwavelength infrared study
of the whole outburst event for the first time.

Triggered by the news on the outburst of OO\,Ser we activated a Target
of Opportunity programme on the {\it Infrared Space Observatory}
\citep[ISO,][]{iso}. The infrared monitoring started 4 months after
the maximum brightness of OO\,Ser, and continued for 20 months. The
ISO-SWS measurements from this programme were published separately by
\citet{larsson}. They found that OO\,Ser changed its infrared fluxes
in the $2-45\,\mu$m range and estimated an extinction of $A_V \approx
20\,$mag from the optical depth of the $10\,\mu$m silicate absorption
feature. In an independent programme the Serpens core was surveyed by the
ISOCAM instrument providing $6.7$ and $14.3\,\mu$m photometry on
OO\,Ser for a single epoch \citep{kaas2004}.

In this paper we analyse our ISOPHOT and ISOCAM observations from the
monitoring programme. In addition to ISO measurements, we observed
OO\,Ser from the ground at $2.2\,\mu$m in 2004 and 2006 as well as at
$12\,\mu$m in 2004. We complemented this database with archival
Spitzer data also from 2004, as well as with previously published
measurements on OO\,Ser from the literature.


\section{Observations and data reduction}
\label{sec:obs}

\begin{table*}
\centering          
\begin{tabular}{lcc@{}ccccccc}
\hline
Date         & ISO\_id  & $\lambda$ [$\mu$m] & Aper./pix. size [$\arcsec$] & Obs. mode & Map size & $\Delta$ [$\arcsec$] & Flux [Jy]  & Abs.~unc. & Rel.~unc. \\ \hline
1996 Feb 28  & 10300901 & $3.6$              & $13.8$           &   Mapping & $3 \times 3$   & $50 \times 50$   & 0.141 & 40\% & 11\% \\
             & 10300901 & $4.8$              & $18$             &   Mapping & $3 \times 3$   & $50 \times 50$   & 1.08  & 40\% & 1\% \\
             & 10300901 & $12$               & $52$             &   Mapping & $3 \times 3$   & $50 \times 50$   & 4.43  & 40\% & 25\%  \\
             & 10300902 & $25$               & $52$             &   Mapping & $3 \times 3$   & $50 \times 50$   & 29.5  & 10\% & 10\% \\ \hline

1996 Apr 14  & 14901404 & $3.6$              & $13.8$           &   Mapping & $3 \times 1$   & $55$             & 0.125 & 40\% & 10\% \\
             & 14901404 & $4.8$              & $18$             &   Mapping & $3 \times 1$   & $55$	        & 0.99  & 40\% & 3\% \\
             & 14901404 & $12$               & $52$             &   Mapping & $3 \times 1$   & $55$	        & 4.07  & 40\% & 3\% \\
             & 14901405 & $25$               & $52$             &   Mapping & $3 \times 1$   & $55$	        & 39.5  & 10\% & 3\% \\ \hline

1996 Sep 01  & 29000208 & $3.6$              & $13.8$           &   Mapping & $3 \times 1$   & $55$             & 0.090 & 40\% & 8\% \\
             & 29000208 & $4.8$              & $18$             &   Mapping & $3 \times 1$   & $55$             & 0.69  & 40\% & 4\% \\
             & 29000208 & $12$               & $52$             &   Mapping & $3 \times 1$   & $55$	        & 3.18  & 40\% & 5\% \\
             & 29000208 & $15$               & $52$             &   Mapping & $3 \times 1$   & $55$	        & 5.10  & 40\% & 14\% \\
             & 29000210 & $60$               & $43 \times 43$   &   Mapping & $9 \times 3$   & $43 \times 43$   & 100   & 17\%  & - \\
             & 29000210 & $100$              & $43 \times 43$   &   Mapping & $9 \times 3$   & $43 \times 43$   & 116   & 19\% & - \\
             & 29000212 & $200$              & $89 \times 89$   & PHT32 map & $27 \times 20$ & $30 \times 30$   & 203   & 11\%  & - \\
             & 29000213 & $2-12$             & $24 \times 24$   &   SP      & $3 \times 1$   & $45$	        & -     & -    & - \\ \hline

1996 Oct 06  & 32500703 & $170$              & $89 \times 89$   & PHT32 map & $5 \times 8$   & $90 \times 46$   & 162   & 15\% & - \\ \hline

1996 Oct 24  & 34300314 & $3.6$              & $13.8$           &   Mapping & $3 \times 1$   & $55$	        & 0.073 & 40\% & - \\
             & 34300314 & $4.8$              & $18$             &   Mapping & $3 \times 1$   & $55$	        & 0.62  & 40\% & - \\
             & 34300314 & $12$               & $52$             &   Mapping & $3 \times 1$   & $55$	        & 3.73  & 40\% & - \\
             & 34300314 & $15$               & $52$             &   Mapping & $3 \times 1$   & $55$             & 4.89  & 40\% & - \\
             & 34300315 & $25$               & $52$             &   Mapping & $3 \times 1$   & $55$             & 21.1  & 10\% & - \\
             & 34300316 & $60$               & $43 \times 43$   &   Mapping & $9 \times 3$   & $43 \times 43$   & 101   & 16\%  & - \\ 
             & 34300316 & $100$              & $43 \times 43$   &   Mapping & $9 \times 3$   & $43 \times 43$   & 122   & 21\% & - \\ 
             & 34300318 & $2-12$             & $24 \times 24$   &   SP      & $3 \times 1$   & $45$             & -     & -    & - \\ \hline

1997 Mar 08  & 47800219 & $3.6$              & $13.8$           &   Mapping & $3 \times 1$   & $55$             & 0.076 & 40\% & 10\% \\
             & 47800219 & $4.8$              & $18$             &   Mapping & $3 \times 1$   & $55$             & 0.53  & 40\% & 5\% \\
             & 47800219 & $12$               & $52$             &   Mapping & $3 \times 1$   & $55$             & 2.98  & 40\% & 2\% \\
             & 47800219 & $15$               & $52$             &   Mapping & $3 \times 1$   & $55$             & 4.50  & 40\% & 3\% \\
             & 47800220 & $25$               & $52$             &   Mapping & $3 \times 1$   & $55$             & 18.5  & 10\% & 2\% \\
             & 47800221 & $60$               & $43 \times 43$   &   Mapping & $9 \times 3$   & $43 \times 43$   & 88.2  & 16\%  & - \\
             & 47800221 & $100$              & $43 \times 43$   &   Mapping & $9 \times 3$   & $43 \times 43$   & 124   & 20\% & - \\
             & 47800223 & $2-12$             & $24 \times 24$   &   SP      & $3 \times 1$   & $45$             & -     & -    & - \\ \hline

1997 Apr 12  & 51301125 & $3.6$              & $13.8$           &   Mapping & $3 \times 1$   & $55$             & 0.060 & 40\% & 8\% \\
             & 51301125 & $4.8$              & $18$             &   Mapping & $3 \times 1$   & $55$             & 0.61  & 40\% & 4\% \\
             & 51301125 & $12$               & $52$             &   Mapping & $3 \times 1$   & $55$             & 3.25  & 40\% & 6\% \\
             & 51301125 & $15$               & $52$             &   Mapping & $3 \times 1$   & $55$             & 4.20  & 40\% & 4\% \\
             & 51301126 & $25$               & $52$             &   Mapping & $3 \times 1$   & $55$             & 18.9  & 10\% & 2\% \\
             & 51301127 & $60$               & $43 \times 43$   &   Mapping & $9 \times 3$   & $43 \times 43$   & 85.3  & 15\%  & - \\
             & 51301127 & $100$              & $43 \times 43$   &   Mapping & $9 \times 3$   & $43 \times 43$   & 121   & 19\% & - \\
             & 51301124 & $200$              & $89 \times 89$   & PHT32 map & $27 \times 20$ & $30 \times 30$   & 233   & 13\%  & - \\
             & 51301129 & $2-12$             & $24 \times 24$   &   SP      & $3 \times 1$   & $45$             & -     & -    & - \\ \hline

1997 Sep 22  & 67601730/31 & $4.8$           & $18$             &   ON/OFF  &   -            & $55$             & 0.44  & 40\% & - \\
             & 67601730/31 & $12$            & $52$             &   ON/OFF  &   -            & $55$             & 2.03  & 40\% & - \\
             & 67601730/31 & $15$            & $52$             &   ON/OFF  &   -            & $55$             & 3.76  & 40\% & - \\
             & 67601730/31 & $25$            & $52$             &   ON/OFF  &   -            & $55$             & 17.1  & 10\% & - \\
             & 67601732    & $60$            & $43 \times 43$   &   Mapping & $9 \times 3$   & $43 \times 43$   & 82.9  & 16\%  & - \\
             & 67601732    & $100$           & $43 \times 43$   &   Mapping & $9 \times 3$   & $43 \times 43$   & 118   & 20\% & - \\
             & 67601734    & $2-12$          & $24 \times 24$   &   SP      & $3 \times 1$   & $45$             & -     & -    & - \\
             & 67601735    & $6.7$           & $1.5 \times 1.5$ &   ISOCAM  & $44 \times 44$ & $1.5 \times 1.5$ & 0.712 & 3.3\% & - \\ 
             & 67601735    & $14.3$          & $1.5 \times 1.5$ &   ISOCAM  & $44 \times 44$ & $1.5 \times 1.5$ & 3.53  & 4.8\% & - \\ \hline
\end{tabular}
\caption{Log of ISOPHOT and ISOCAM observations. SP stands for
  spectrophotometry. ``Map size'' indicates the sizes of the final
  maps. $\Delta$ denotes the increment between adjacent pixel
  positions in the map (mapping) or the separation between source and
  background positions (ON/OFF). At certain wavelengths the beam
  contained nearby sources; for a detailed discussion see
  Sec.~\ref{sec:conf}. All fluxes are colour corrected. The last two
  columns give the uncertainties of the absolute and the relative flux
  calibration (Sec.~\ref{sec:isophot}).}
\label{tab:ISOobs}
\end{table*}

\subsection{ISOPHOT observations} 

ISOPHOT, the photometer on-board ISO \citep{isophot}, carried out
multi-filter photometry with 9 different filters in the
$3.6{-}200\,\mu$m wavelength range and spectrophotometry in the
$2.47{-}11.62\,\mu$m range, at 8 different epochs between 1996
February and 1997 September. Table \ref{tab:ISOobs} shows the log of
the observations. In most cases small raster maps were obtained except
in 1997 September, when at $25\,\mu$m and shortwards the source and
background positions were observed separately. The typical integration
time was $64\,$s. Aperture sizes varied according to the filters:
$13\farcs8$ at $3.6\,\mu$m, $18''$ at $4.8\,\mu$m, $52\arcsec$ at
$12$, $15$ and $25\,\mu$m. At $60$ and $100\,\mu$m the C100 camera
($3{\times}3$ pixel, $43\arcsec{\times}43\arcsec$ per pixel), while at
$170$ and $200\,\mu$m the C200 camera ($2{\times}2$ pixel,
$89\arcsec{\times}89\arcsec$ per pixel) was utilised. Some
far-infrared observations were performed in the PHT\,32 mode, which
provided higher spatial resolution \citep[for a description of this
mode see ][]{pht32}.

\subsection{ISOPHOT data processing}
\label{sec:isophot}

\paragraph{Standard processing and absolute flux calibration.} The
data reduction was performed using the ISOPHOT Interactive Analysis
Software Package V10.0 \citep[PIA,][]{pia}. The integration ramps were
corrected for non-linearities. Cosmic particle hits were removed using
the two thresholds deglitching method, and signal values were derived
by fitting a first order polynomial to each ramp. The signals were
transformed to a standard reset interval, then an orbital dependent
dark current was subtracted and cosmic ray hits were again checked. In
case the signal did not fully stabilise during the measurement time
due to detector transients, only the last part of the data stream was
used. This was found mainly in observations with the $12$ and
$25\,\mu$m filters, while at other wavelengths the measurements showed
sufficient stability. The absolute flux calibration of measurements at
$25\,\mu$m and shortwards was done by adopting the default
responsivity of the detector. The ISOPHOT Handbook
\citep{isophot_handbook} gives a typical absolute flux uncertainty of
40\% at $3.6$, $4.8$, $12$ and $15\,\mu$m, and 10\% at $25\,\mu$m for
ON/OFF staring measurements. Since our small raster maps enabled
better background subtraction than the ON/OFF mode, the error of our
measurements is probably smaller. Nevertheless, we adopted the
abovementioned values as conservative estimates for the absolute flux
uncertainty (see Col.~9 of Table~\ref{tab:ISOobs}).

\paragraph{Relative flux determination.} Due to the monitoring
strategy, the observations were carried out with identical instrument
setup at most epochs. This enabled us to determine relative flux
variations at a certain wavelength more accurately than the standard
processing by applying a new method developed at Konkoly Observatory
especially for the fine relative calibration of ISOPHOT data
\citep{juhasz}. The algorithm compares the detector transient curves
in a measurement sequence with the corresponding curves of a reference
day (in our case 1996 October 24) and determines a scaling factor
between the signals of the two epochs. Evaluating all epochs, one
obtains a light curve normalized to the reference day. The final
fluxes are computed as the product of the scaling factors and the
absolute flux level of the reference day. The method provides relative
uncertainties, which correspond to the formal errors of the scaling
factors. In our case, the relative uncertainties were usually around
or below 10\% (see Col.~10 of Table~\ref{tab:ISOobs}). The absolute
flux level of the whole light curve is determined by the absolute
calibration of the reference flux.

\paragraph{Far-infrared maps.} The far-infrared observations
at $60$ and $100\,\mu$m were processed with PIA in a standard
way. PHT32 observations at $170$ and $200\,\mu$m were reduced using a
dedicated software package (P32TOOLS) developed at MPI Kernphysik in
Heidelberg \citep{pht32}. This tool provides adequate correction for
transients in PHT32 measurements. Absolute calibration was done by
comparing the source flux with the on-board fine calibration
source. At $60$ and $100\,\mu$m each $9\,{\times}\,3$ map was
flat-fielded using the first raster step ($3\,{\times}\,3$ pixel) as
the sky flat position. Then the sum of two point spread functions,
centred on the positions of OO\,Ser and a nearby submillimetre source,
\object{SMM\,1} (see Sec.~\ref{sec:conf}), was fitted to the
brightness distribution on the map using the ISOPHOT measured
footprint maps. A similar fitting procedure was applied to the PHT32
maps. The errors given in Table~\ref{tab:ISOobs} represent the
quadratic sum of the formal uncertainties of the fits (in the range of
3--15\%) and the photometric calibration uncertainty of the detector
given in the ISOPHOT Handbook \citep{isophot_handbook}. We note that
with this technique OO\,Ser could be separated from SMM\,1 but not
from other closer nearby sources, which may contaminate the flux of
OO\,Ser. $9\,{\times}\,3$ maps at $60$ and $100\,\mu$m were obtained
at 7 different epochs, but the first two measurements (in 1996
February and April) were executed at late orbital phases during ISO's
orbit. This caused a large uncertainty in these measurements, thus we
decided not to present them. In the PHT32 oversampled maps more than
one detector pixel observed OO\,Ser; their independent photometric
results were combined with a robust averaging technique described by
\citet{fuors}.

\paragraph{ISOPHOT-S.} Spectrophotometric observations, also in the
form of $3\,{\times}\,1$ small raster maps centred on OO\,Ser, were
obtained with the ISOPHOT-S subinstrument. The processing of these
data deviated from the standard scheme implemented in PIA. ISOPHOT-S
has a successful dynamic calibration for staring observations but in
PIA this method is not applicable for rasters, leading to uncorrected
transients and consequently reduced photometric accuracy. Since the
contrast between OO\,Ser and the background level is relatively low,
we treated these rasters as long staring observations, and applied the
dynamic calibration by modifying some PIA routines. The resulting
photometric uncertainties are in the range of 5--10\%.

\paragraph{Colour corrections.} Colour corrections were applied
to each broadband photometric measurement by convolving the observed
spectral energy distribution (SED) at a certain epoch with the ISOPHOT
filter profile in an iterative way. The result of the ISOPHOT
photometry is presented in Table~\ref{tab:ISOobs}.

\subsection{ISOCAM observations and data processing}

The $6.7$ and $14.3\,\mu$m photometry was obtained with the ISOCAM
instrument \citep{isocam} on 1997 September 22 (see
Table~\ref{tab:ISOobs}). The data were reduced with the CAM
Interactive Analysis Software V5.0 \citep[CIA,][]{cia}. A dark current
correction was applied following the `VilSpa' method. Glitches were
removed using the `multiresolution median transform'. This efficiently
removes glitches based on the fact that glitches in general are much
shorter than the signature of a real source. The data were stabilised
using the Fouks-Schubert model. Next the individual frames were
averaged to four mean images, one for each sky position. After this, a
second deglitching was applied based on the overlapping projected sky
positions of these mean images. Finally, the images were flat-fielded
and combined into the final mosaic. The pixel values were converted to
mJy/arcsec$^2$ using the tabulated conversion factors available in
CIA.

Photometry of the sources was obtained using the IRAF tool `xphot'. We
used a small aperture photometry of $4\farcs5$ and $6\arcsec$ radius
for the $6.7$ and $14.3\,\mu$m mosaic, respectively. These values were
then corrected for the flux falling in the wings of the point spread
function outside the chosen aperture. The small aperture was selected
because of a nearby source (\object{V370\,Ser} at a distance of
${\sim}\,11\arcsec$) which otherwise would have contributed to the
measured flux. The results are shown in Table~\ref{tab:ISOobs}. The
insignificant background level of 1 and 3 mJy/arcsec$^2$ does not
contribute to the measurement error, which should be dominated by the
absolute flux calibration uncertainty of 3.3\% at $6.7\,\mu$m and
4.8\% at $14.3\,\mu$m \citep{2003ESASP1262.....B}.

\subsection{Near-infrared observations and data processing}

K$_S$-band images were obtained by the LIRIS instrument on 2004 June
11 (as part of the LIRIS Guaranteed Time programme) and on 2006 May
6. LIRIS is an infrared camera/spectrograph built at the Instituto de
Astrof\'\i{}sica de Canarias \citep{ap, man}, and is mounted on the
4.2\,m William Herschel Telescope at the Observatorio del Roque de los
Muchachos (Canary Islands). The detector used is a
$1024\,{\times}\,1024$ HAWAII--1 detector \citep{hawaii}, which
provides a plate scale of $0\farcs250$/pixel and a total area of
$4\farcm27\,{\times}\,4\farcm27$. In 2004 a three point dither pattern
was used with a total exposure time of 140\,s, while in 2006 a
five-point dither pattern was used and the total exposure time was
250\,s. The images were reduced using the IRAF package `liris\_ql'
developed by the LIRIS team within the IRAF environment. The reduction
steps consist of flat-fielding, sky subtraction and image
co-addition.

\begin{figure*}
\centering
\includegraphics[angle=90,width=181.mm]{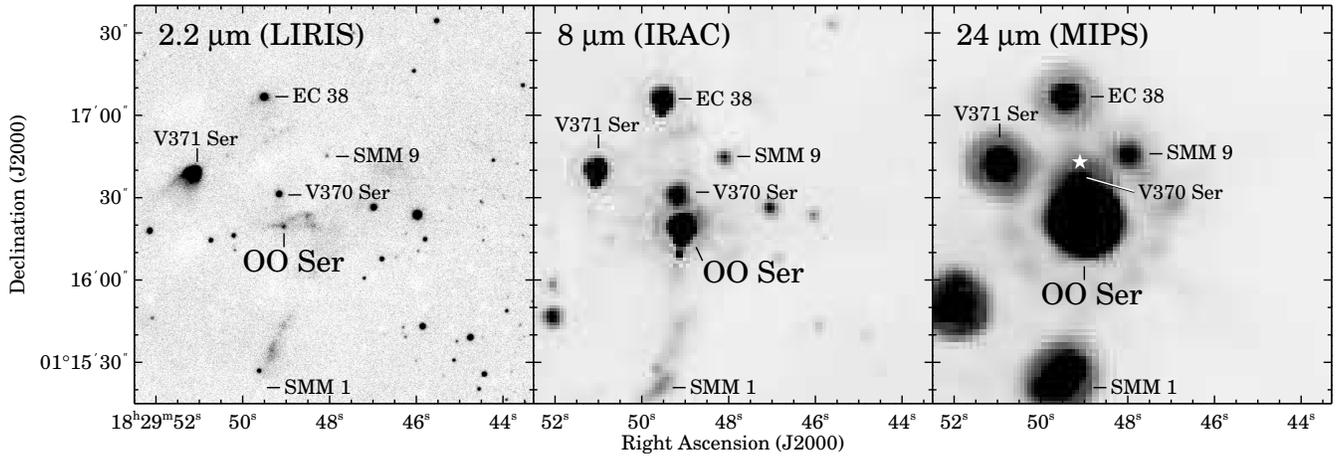}
\caption{OO\,Ser and its surroundings at different infrared
  wavelengths. V370\,Ser (EC\,37), V371\,Ser (EC\,53), EC\,38, SMM\,9
  and SMM\,1 are also marked. In the middle panel, the spots below the
  brightest sources are instrumental artifacts ('bandwidth
  effects'). The white star in the right panel indicates the position
  of IRAS18272+0114 as given in the IRAS Catalogue of Point Sources.}
\label{fig:liris}
\end{figure*}

In Fig.~\ref{fig:liris} (left) a part of our K$_S$-band LIRIS image
from 2004 can be seen showing the surroundings of OO\,Ser. The
eruptive star was clearly detected in K$_S$ and a faint nebulosity
around the star can be seen as well (see also
Sec.~\ref{sec:morph}). Because of this nebulosity, photometry should
be done carefully. In order to be able to compare our measurements
with those of \citet{hodapp99}, we used an aperture diameter of
$11{\farcs}3$, the same as \citet{hodapp99}. Conversion from
instrumental magnitudes to real magnitudes was done using the 2MASS
J--K$_S$ colour versus $\Delta$K$_S$ (the difference between the
instrumental and the 2MASS K$_S$ magnitudes) relationship for 20
comparison stars in the field. The resulting values are K$_S =
14.3\,{\pm}\,0.2$ mag in 2004 and K$_S = 14.1\,{\pm}\,0.2$ mag in
2006. We note that the colours of the 20 comparison stars used in this
calibration procedure covered a large enough range to include also the
colour of OO\,Ser itself. Similar photometry was derived for some
nearby sources (similarly to \citet{hodapp99}, we used a larger,
$11\farcs3$ diameter aperture for the two extended objects: OO\,Ser
and V371\,Ser, and a smaller $1\farcs9$ diameter aperture with
the corresponding aperture correction for the point-like objects:
V370\,Ser, EC\,38 and SMM\,9); the results can be seen in
Table~\ref{tab:2004}.

\begin{figure}
\centering
\includegraphics[angle=90,width=\columnwidth]{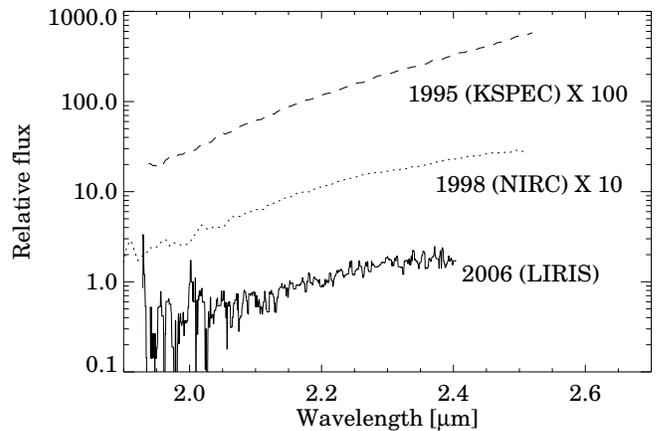}
\caption{Normalised K-band spectra of OO\,Ser. {\it Continuous line:}
this work, {\it dashed and dotted lines:} \citet{hodapp99}. For
clarity, we multiplied the spectrum from 1998 by 10 and the spectrum
from 1995 by 100.}
\label{fig:spec}
\end{figure}

A K-band spectrum of OO\,Ser was obtained using LIRIS on 2006 May
6. The observation was performed following an ABBA telescope nodding
pattern. The total exposure time was 2400\,s, split in 4 individual
exposures of 600\,s. In order to reduce the readout noise, the
measurements were done using multiple correlated readout mode, with 4
readouts before and after the integration. We used a slit width of
1\arcsec{} and a medium resolution sapphire grism which yielded a
spectral resolution of 2500. The wavelength calibration was provided
by observations of an Argon lamp available in the calibration unit at
the A\&G box of the telescope. In order to obtain the telluric
correction, the nearby A0V star HIP~90123 was observed with the same
configuration as the object. The data were reduced and calibrated
using the package `liris$\_$ql'. Consecutive pairs of AB
two--dimensional images were subtracted to remove the sky background,
then the resulting images were wavelength calibrated and flat-fielded
before registering and coadding all frames to provide the final
combined spectrum. A one dimensional spectrum was extracted with the
IRAF `apall' task. The extracted spectrum was divided by a composite
to eliminate telluric contamination. This composite spectrum was
generated from the observed spectrum of the calibration star, divided
by a stellar model and convolved to our spectral resolution.  The
resulting normalised K-band spectrum is shown in Fig.~\ref{fig:spec}.

\subsection{Mid-infrared observations and data processing}

OO\,Ser was imaged with TIMMI2 mounted on the ESO 3.6\,m telescope at
La Silla on 2004 October 21, under clear and stable conditions. The
N11.9-OCLI filter was used which has a central wavelength of
$11.66\,\mu$m and a FWHM of $1.16\,\mu$m. The total integration time
was 4 minutes. Both chopping and nodding amplitudes were $10\arcsec$,
and the $0\farcs2$ pixel scale of the $240\,{\times}\,320$ Raytheon
detector was set. This resulted in a chop--nod corrected image with
two negative and two positive beams captured on the detector. The four
beams were used for independent determination of the source flux and
error. The resulting fluxes are 0.64 Jy for OO\,Ser and 0.18 Jy for
V370\,Ser (a nearby young star also visible in the field, see also
Table~\ref{tab:2004}). The flux calibrator was the photometric
standard \object{HD\,96171} which was observed with the same
set-up. Other photometric standard star observations of the same night
were inspected to estimate the total absolute flux uncertainty of
${\sim}15\%$.

\subsection{Spitzer observations and data processing}

OO\,Ser was measured with the IRAC and MIPS instruments of the Spitzer
Space Telescope on 2004 April 4 and 5, respectively, as part of the
legacy programme `c2d' (PI: Neal J.~Evans II). The third data
release of this programme containing enhanced data products and
catalogues can be downloaded from the Spitzer
website\footnote{http://ssc.spitzer.caltech.edu/legacy/all.html}.
As an example we plotted parts of the $8\,\mu$m IRAC and the
$24\,\mu$m MIPS maps in Fig.~\ref{fig:liris}.

The `c2d' catalogue contains IRAC fluxes for OO\,Ser, but it is
treated as a point source. The nebulosity seen in the K-band images,
however, are still visible at 3.6 and 4.5$\,\mu$m too. In order to be
able to compare the IRAC fluxes with the ISO measurements, we decided
to extract fluxes from the IRAC images for OO\,Ser using the same
apertures as ISOPHOT used (diameter of $13\farcs8$ at $3.6\,\mu$m and
$18"$ at $\ge4.5\,\mu$m)

The current version of the 'c2d' catalogue does not contain MIPS
fluxes yet. In order to obtain photometry in the MIPS bands, we
downloaded the enhanced MIPS images. At $24\,\mu$m we selected 7
isolated stars to construct the point spread function. Since
OO\,Ser was saturated, this profile was then fitted to the
non-saturated wings of OO\,Ser. At $70\,\mu$m OO\,Ser and a nearby
young star, V370\,Ser could not be fully separated. Thus an aperture
of 20$''$ was utilized, which included both objects. The measured flux
was then distributed between OO\,Ser and V370\,Ser in the same ratio
as their peak brightnesses (measured in a 5$''$ aperture at the
position of the two sources). We also extracted $70\,\mu$m photometry
for some other sources in the vicinity of OO\,Ser, since they possibly
cause source confusion in IRAS and ISO far-infrared measurements. In
these cases, a 20$''$ aperture was utilized with a fixed 0.142 Jy
background.

Colour corrections were applied to each measurement for each source by
convolving the observed SED with the IRAC and MIPS filter profiles in
an iterative way. The resulting fluxes and estimated uncertainties can
be seen in Table~\ref{tab:2004}.

\begin{table*}
\begin{minipage}[t]{\textwidth}
\centering
\renewcommand{\footnoterule}{}  
\renewcommand{\thefootnote}{\thempfootnote}
\begin{tabular}{cccccccc}
\hline
$\lambda$ [$\mu$m] & Date & Instrument       &  OO\,Ser                   &        V370\,Ser      &         V371\,Ser         &         EC\,38            &         SMM\,9             \\ 
\hline
2.2                & 2004 Jun 11 & LIRIS     & 1.16 $\pm$ 0.20          & 2.25 $\pm$ 0.39       & 3.76 $\pm$ 0.57           & 5.64 $\pm$ 0.95           & 0.13 $\pm$ 0.02          \\
2.2                & 2006 May 6  & LIRIS     & 1.54 $\pm$ 0.26          & 2.85 $\pm$ 0.44       & 1.60 $\pm$ 0.22           & 4.82 $\pm$ 0.70           & 0.15 $\pm$ 0.02          \\ \hline

3.6                & 2004 Apr 4 & IRAC       & 7 $\pm$ 1                  & 17 $\pm$ 1            & 10 $\pm$  1               & 43.4 $\pm$ 0.4            & 1.9 $\pm$  0.2             \\ 
4.5                & 2004 Apr 4 & IRAC       & 82 $\pm$  2                & 38 $\pm$ 1            & 17 $\pm$  1               & 101 $\pm$  1              & 3.1 $\pm$  0.2             \\ 
5.8                & 2004 Apr 4 & IRAC       & 81 $\pm$  4                & 63 $\pm$ 1            & 140 $\pm$ 2               & 139 $\pm$  1              & 12.4 $\pm$ 0.7             \\ 
8.0                & 2004 Apr 4 & IRAC       & 570 $\pm$ 20               & 86 $\pm$ 2            & 190 $\pm$ 6               & 162 $\pm$  2              & 14.4 $\pm$ 0.4             \\ \hline

12                 & 2004 Oct 21 & TIMMI2    & 640 $\pm$ 100              & 180 $\pm$ 30          & -                         & -                         & -                          \\ \hline

24                 & 2004 Apr 5 & MIPS       & 13~300 $\pm$ 1~800         & 1~560 $\pm$ 60        & 1~010 $\pm$ 20            & 530 $\pm$ 10              & 227 $\pm$ 10               \\
70                 & 2004 Apr 5 & MIPS       & 14~000 $\pm$ 600           & 900 $\pm$ 600         & 7~900 $\pm$ 220          & 5~200 $\pm$ 1~400         & 13~400 $\pm$ 4~000         \\
\hline
\end{tabular}
\end{minipage}
\caption{Log of observations from 2004--2006. All fluxes are presented
  in mJy. Spitzer fluxes are from the third delivery of data from the
  `c2d' legacy project, except the $70\,\mu$m data of all sources and
  the $3.6$, $4.5$, $5.8$, $8$ and $24\,\mu$m data of OO\,Ser, which
  were extracted by us from MIPS and IRAC images improved and
  published by the `c2d' legacy team. All Spitzer fluxes are colour
  corrected.}
\label{tab:2004}
\end{table*}

\subsection{Source confusion}
\label{sec:conf}

Fig.~\ref{fig:liris} shows that several infrared and submillimetre
sources are present in the vicinity of OO\,Ser: V370\,Ser (also known
as EC\,37), \object{V371\,Ser} (also known as EC\,53 or SMM\,5),
\object{EC\,38}, \object{SMM\,9} and SMM\,1. At shorter wavelengths
($2.2$, $3.6$, $4.5$ and $5.8\,\mu$m), an extended nebulosity around
OO\,Ser can also be seen. Thus, when comparing the fluxes of OO\,Ser
measured with different instruments one has to keep in mind that--to
some extent--the nebulosity and some of the abovementioned sources may
contribute to the observed flux at a particular wavelength. At
$2.2\,\mu$m and with the ISOPHOT/ISOCAM at $3.6$, $4.8$, $6.7$ and
$14.3\,\mu$m the beam included OO\,Ser only. At $12$, $15$ and
$25\,\mu$m the ISOPHOT beams included OO\,Ser and V370\,Ser. At $60$,
$100$, $170$ and $200\,\mu$m the fluxes extracted for OO\,Ser include
also contributions from V370\,Ser, V371\,Ser, EC\,38 and
SMM\,9. Recent infrared photometry for these sources can be seen in
Table~\ref{tab:2004}. It should be noted that fluxes presented in
Table~\ref{tab:ISOobs} contain the contributions of nearby sources as
discussed above. In Sec.~\ref{sec:light} we give a detailed
description of how we corrected the ISOPHOT measurements for the
effects of source confusion.


\section{Results}
\label{sec:results}

\subsection{Morphology of the nebula}
\label{sec:morph}

According to \citet{hodapp99} before the outburst a triangle-shaped
nebula west of OO\,Ser and a small elongated nebula east of OO\,Ser
could be seen. During the outburst these -- presumably reflection --
nebulae became much brighter. In this phase \citet{hodapp96} observed
OO\,Ser in the L$^{\prime}$ and M$^{\prime}$ bands, as well as at
$11.7$ and $20.6\,\mu$m. They found that the nebulosity can be seen in
the L$^{\prime}$ but not at longer wavelengths. Our recent K$_S$-band
images reveal that the nebulae still exist and look very similar to
the preoutburst image of \citet{hodapp99} (see his Fig.~1 left). The
nebulosity around OO\,Ser is also visible in the Spitzer/IRAC maps
from 2004 at $3.6, 4.5$ and $5.8\,\mu$m and some extended
emission can be suspected even at $8\,\mu$m.

\subsection{Preoutburst fluxes}
\label{sec:pre}

In order to study the consequences of the outburst in the whole
infrared regime, one has to compile first the SED of OO\,Ser in the
quiescent phase, i.e. estimate the preoutburst fluxes. At $2.2\,\mu$m
there exists a preoutburst measurement from 1994 \citep{hodapp96}. At
longer wavelengths, only the IRAS measurements are available from
1983. Analysing high-resolution IRAS maps, \citet{barsony} derived
fluxes of 0.63, 4.5, 24 and 111 Jy at 12, 25, 60 and 100$\,\mu$m
respectively. The authors claim that these values represent the total
fluxes from a region encompassing three confused sources: OO\,Ser,
V371\,Ser and SMM\,9, and they give the one third of the
abovementioned values as upper limits for the brightness of
OO\,Ser. The position of IRAS18272+0114 from the IRAS Catalogue of
Point Sources (marked by a white star in Fig.~\ref{fig:liris} right)
is located halfway between the sources, which may indicate that there
was no dominant source, but all sources had comparable contributions
to the IRAS flux.

With the help of Spitzer/MIPS measurements, it is possible to check
the validity of the assumption of \citet{barsony} via estimating
the preoutburst fluxes of OO\,Ser at $25$ and $60\,\mu$m. Assuming
that EC\,38 and SMM\,9 have non-variable far-infrared fluxes, we
subtracted the contribution of these sources from the IRAS values
cited above. Another nearby source, V371\,Ser, exhibits
near-infrared variability of ${\approx}\,1.5\,$mag and shares many
characteristics with EXors \citep{hodapp99}. At far-infrared
wavelengths, however, eruptive young stars do not typically show
significant flux changes \citep{fuors}. Thus we assumed that V371\,Ser
is also non-variable at $25$ and $60\,\mu$m, at least within our
measurement uncertainties, and we subtracted its flux from the IRAS
values. In practice, we estimated $25\,\mu$m fluxes for these nearby
sources via interpolating from the $24\,\mu$m MIPS values in
Table~\ref{tab:2004}, and subtracted the sum of these ($3.6$\,Jy) from
the value given by \citet{barsony} ($4.5$\,Jy). The result is
$0.9$\,Jy, which is indeed of the order of one third of $4.5$\,Jy. The
result at $60\,\mu$m is also roughly consistent with the one third
value. Due to the lack of recent $12$ or $100\,\mu$m data for all
nearby sources, the same test cannot be done at these
wavelengths. Therefore, for homogeneity, at all four IRAS wavelengths
we adopted as preoutburst fluxes the one third values within a factor
of 2 uncertainty: $0.21^{+0.21}_{-0.10}\,$Jy at $12\,\mu$m,
$1.5^{+1.5}_{-0.8}\,$Jy at $25\,\mu$m, $8^{+8}_{-4}\,$Jy at
$60\,\mu$m, $37^{+37}_{-19}\,$Jy at $100\,\mu$m. We note that OO\,Ser
is the first eruptive young star where preoutburst fluxes are
available in the whole infrared wavelength regime. Similar preoutburst
data can be found only for one other young eruptive star, V1647\,Ori
\citep{v1647ir}.

\subsection{Spectral energy distribution}
\label{sec:sed}

\begin{figure}
\centering
\includegraphics[angle=90,width=\columnwidth]{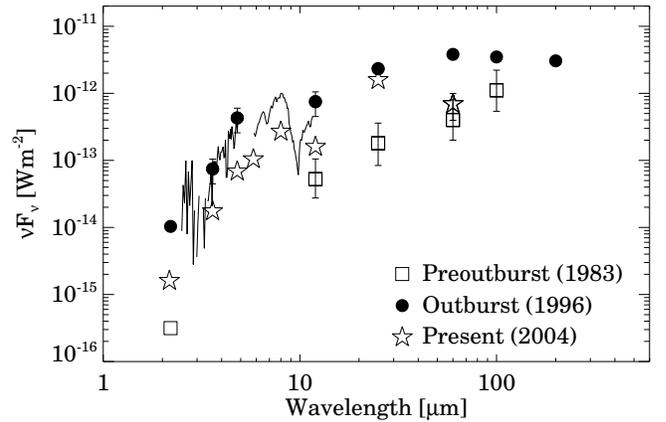}
\caption{Spectral energy distribution of OO\,Ser. {\it Squares}:
 preoutburst fluxes measured with IRAS in 1983 \citep{barsony}, and
 K-band photometry from 1994 August, \citep{hodapp96}; {\it Dots and
 line}: outburst fluxes from 1996 September, measured with ISOPHOT;
 {\it Stars}: current fluxes from 2004, measured with LIRIS,
 TIMMI2 and Spitzer. Error bars smaller than the symbol size are not
 plotted. ISOPHOT beams at $100$ and $200\,\mu$m contained nearby
 sources; for a detailed discussion see Sec.~\ref{sec:conf} and
 \ref{sec:sed}.}
\label{fig:sed}
\end{figure}

\begin{figure}
\centering
\includegraphics[angle=90,width=\columnwidth]{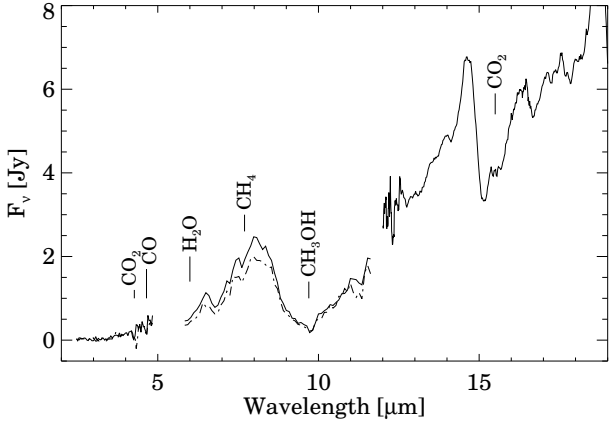}
\caption{Ice features in the spectrum of OO\,Ser. Data below
$12\,\mu$m were taken by ISOPHOT-S. Solid line indicates data from
1996 October, dashed dot line from 1997 April. Data above $12\,\mu$m
were taken by the ISO-SWS instrument \citep[highly processed data products
from ][]{hpdp}. In order to obtain high signal-to-noise ratio,
ISO-SWS spectra from 1997 March, April and September were averaged and
a $0.3\,\mu$m wide moving median window was applied.}
\label{fig:ices}
\end{figure}

In Fig.~\ref{fig:sed} the SED of OO\,Ser in the high (outburst) state
from 1996 September is plotted with filled symbols. We attempted to
correct for the effects of source confusion as described in details in
Sec.~\ref{sec:light}. Thus, the resulting SED in Fig.~\ref{fig:sed}
represents the flux from OO\,Ser alone at ${\le}\,60\,\mu$m. Due to
high extinction the star is invisible at optical wavelengths and is
very faint in the near-infrared regime. In the mid-infrared
($3{-}25\,\mu$m) the SED is rising towards longer wavelengths. This
wavelength regime is zoomed in Fig.~\ref{fig:ices}, showing a broad
silicate absorption feature at $9.7\,\mu$m, as well as several ice
features, which also indicate high extinction. Measuring the optical
depth of the $9.7\,\mu$m feature, and converting it to visual
extinction assuming $A_V /\Delta\tau_{9.7} = 18.5$ \citep{draine2003},
we obtained $A_V = 42\,{\pm}\,5\,$mag. This value is twice as large as
the one measured by \citet{larsson}; the difference is probably related
to the low signal-to-noise of his SWS spectra around the $9.7\,\mu$m
absorption feature. The measured high extinction indicates that
OO\,Ser is more deeply embedded than most FUors. At far infrared
wavelengths ($60{-}200\,\mu$m) the SED is flat, but it should be noted
that data points at $100$ and $200\,\mu$m are contaminated by source
confusion (see Sec.~\ref{sec:conf}).

In Fig.~\ref{fig:sed} the preoutburst SED (see Sec.~\ref{sec:pre}) and
the present SED (based on measurements from 2004) are also
plotted. They will be discussed in Sec.~\ref{sec:ini} and
\ref{sec:present}.

\subsection{Light curves}
\label{sec:light}

ISOPHOT, ISOCAM, LIRIS, TIMMI2, Spitzer, IRAS \citep{barsony} and
K-band measurements \citep{hodapp96, hodapp99} were combined to
construct the light curves of OO\,Ser at different wavelengths between
1995 and 2006.

For the subsequent light curve analyses, we attempted to correct for
the effects of source confusion. One possibility would be to smooth
all data to the same resolution, in most cases defined by the ISOPHOT
aperture. In doing so, however, several additional sources would be
included in the beam, falsifying the fading rate calculations for
OO\,Ser. Instead, we decided to use the higher spatial resolution
images to correct for the contribution of additional, unrelated
sources in the large ISOPHOT beams.

At $12\,\mu$m TIMMI2 could resolve OO\,Ser and V370\,Ser. Assuming
that V370\,Ser is not variable at this wavelength, we subtracted its
contribution of 0.18\,Jy (see Table~\ref{tab:2004}) from each ISOPHOT
$12\,\mu$m measurement. The MIPS camera of Spitzer at $24\,\mu$m could
separate OO\,Ser from V370\,Ser giving a flux of 1.56\,Jy for the
latter source, which we interpolated to $25\,\mu$m (1.71\,Jy) and
subtracted from the ISOPHOT points at $25\,\mu$m. At $60\,\mu$m, MIPS
$70\,\mu$m measurements could be utilized. Using the SEDs presented in
Table~\ref{tab:2004}, we interpolated $60\,\mu$m fluxes for V370\,Ser,
V371\,Ser, EC\,38 and SMM\,9, and subtracted the sum of these values
(24\,Jy) from the ISOPHOT $60\,\mu$m data points. Due to the lack of
$100\,\mu$m fluxes for the nearby sources, a similar correction was
not possible in the case of the $100\,\mu$m ISOPHOT light curve.

\begin{figure*}
\sidecaption
\includegraphics[width=132mm]{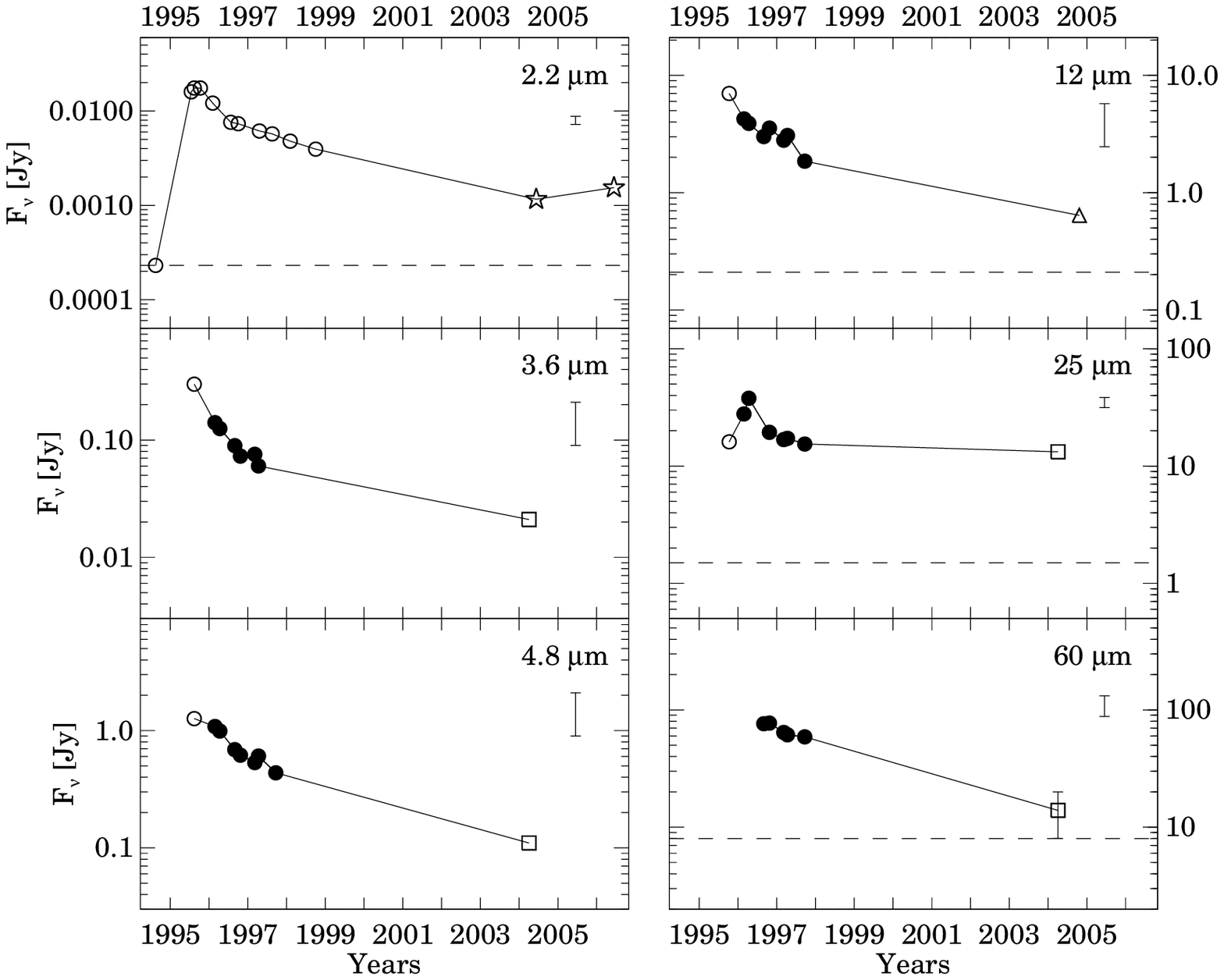}
\caption{Light curves of OO\,Ser at different wavelengths. {\it Open
 circles:} measurements of \citet{hodapp96, hodapp99}; {\it Stars:}
 LIRIS; {\it Filled circles:} ISOPHOT; {\it Triangle:} TIMMI2; {\it
 Squares:} Spitzer; {\it Dashed lines:} preoutburst fluxes from 1994
 at $2.2\,\mu$m and from 1983 at $12$, $25$ and $60\,\mu$m (for
 details see Sec.~\ref{sec:pre}). Tick marks indicate January 1 of the
 corresponding year. Error bars in the upper right corner of each
 panel show the typical uncertainty in the absolute flux level. Due to
 the homogeneous measurement strategy, the uncertainty of the ISOPHOT
 data points relative to each other at a certain wavelength are
 usually better than the absolute errors (see
 Sec.~\ref{sec:isophot}). All fluxes are corrected for source
 confusion (for a detailed description see Sec.~\ref{sec:conf} and
 \ref{sec:light}), thus these light curves show the brightness
 evolution of OO\,Ser alone.}
\label{fig:lightcurve2}
\end{figure*}

In Fig.~\ref{fig:lightcurve2} six representative light curves between
$2.2$ and $60\,\mu$m are shown. All these light curves are corrected
for source confusion and represent the brightness evolution of OO\,Ser
alone. In the following, we describe these light curves in detail.

\subsubsection{The outburst history at $2.2\,\mu$m}
\label{sec:2}

FUor and EXor ourbursts were historically monitored at optical
wavelengths. Since OO\,Ser is invisible in the optical, K$_S$ is the
shortest available band where the outburst could have been followed.
The top left panel of Fig.~\ref{fig:lightcurve2} shows the $2.2\,\mu$m
light curve of OO\,Ser. The star brightened by $4.6$ mag between 1994
August and 1995 July. After reaching peak brightness, it started an
approximately exponential fading with a rate of $1.00$ mag/year in the
first 350 days and $0.34$ mag/year afterwards, as the data between
1995 and 1999 indicate \citep{hodapp99}. The change in fading rate
divides this period of the outburst into a first and a second
part. Our measurements from 2004--2006 prove that the fading
continued, although at a slightly different rate, representing a third
part of the fading. Comparing the preoutburst flux with the new
observations, one can conclude that at $2.2\,\mu$m OO\,Ser is still
above the preoutburst flux level.

In the following, we describe the lightcurves of OO\,Ser at longer
wavelengths, following the abovementioned division: the initial rise
(until peak brightness); the first part of the fading (until mid-1996);
the second part of the fading (until mid-1997); and the third part of
the fading (until 2004).

\subsubsection{The initial rise (until peak brightness)}
\label{sec:ini}

Comparison of the preoutburst fluxes with the SED in outburst
(Fig.~\ref{fig:sed}) shows that the eruption caused brightening in the
whole near- to far-infrared spectrum. Though the fluxes at $100\,\mu$m
are contaminated by nearby sources, there is a flux change at this
wavelength, too. The shape of the SED changed significantly: the
outburst SED is flatter.

At $2.2\,\mu$m the light curve reached its peak in 1995 October. At
$3.6$, $4.8$ and $12\,\mu$m, the exact date of the peak is not known,
but it happened not later than 1995 October, thus it was probably
simultaneous with the K-band peak. At $25\,\mu$m, however, the peak
took place in 1996 April, i.e. some 200 days later than at
$2.2\,\mu$m. The shape of the peak at $25\,\mu$m is also different, it
is broader than at $2.2\,\mu$m, and has a triangle-like shape. At $60$
and $100\,\mu$m, the photometry is more uncertain, thus the peak dates
cannot be determined. Nevertheless, the time-shift at $25\,\mu$m gives
a hint that the peak brightness happened gradually later at longer
wavelengths. To our knowledge, such a time-shift has not been observed
for other eruptive stars.

\begin{figure}
\centering
\includegraphics[angle=0,width=\columnwidth]{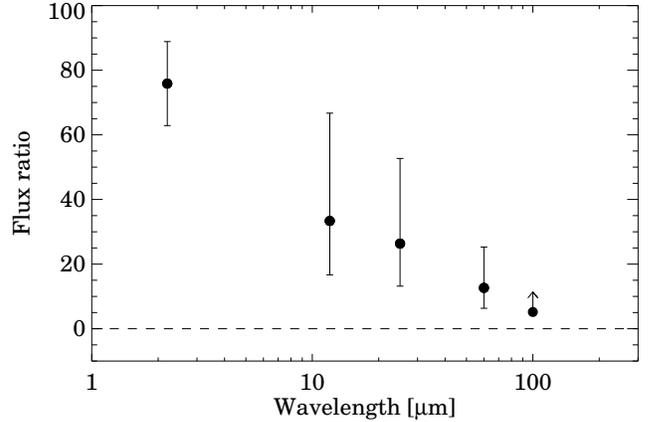}
\caption{Initial rise: flux ratio of peak brightness to preoutburst
  brightness at each wavelength. (See discussion in
  Sec.~\ref{sec:ini}.)}
\label{fig:brightening}
\end{figure}

In order to analyse the wavelength dependence of the brightening, in
Fig.~\ref{fig:brightening} we plotted the ratio of the peak flux to
the preoutburst value at $2.2$, $12$, $25$, $60$ and $100\,\mu$m.
Since at $100\,\mu$m both the preoutburst IRAS flux and the peak
ISOPHOT flux include contributions from several nearby sources, the
ratio derived from these numbers represents a lower limit for the
brightening of OO\,Ser itself. Figure \ref{fig:brightening} suggests
that the amplitude of the flux increase has a characteristic
wavelength dependence: the flux ratio is lower at longer wavelengths,
with an approximately linear dependence on $\log\lambda$. Although the
peak brightness did not occur simultaneously at different wavelengths,
this trend can also be seen if one compares in Fig.~\ref{fig:sed} the
preoutburst SED with the SED from 1996 September (close to peak
brightness). We note that this result is different from the case of
another young eruptive star, V1647\,Ori, where the initial rise was
practically wavelength-independent \citep{v1647ir,muz}.

\begin{figure}
\centering
\includegraphics[angle=0,width=\columnwidth]{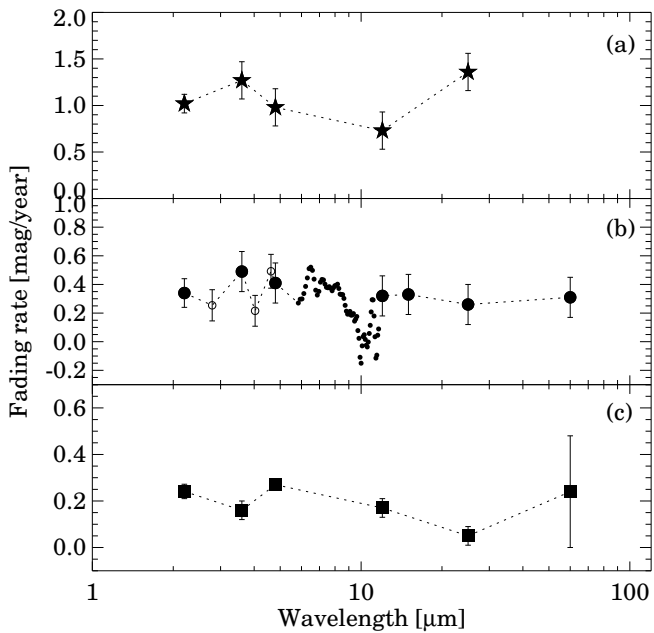}
\caption{Fading rates of OO\,Ser at different wavelengths, {\it (a)}
  between 1996 April and 1996 September, {\it (b)} between 1996 October
  and 1997 March, {\it (c)} between 1997 April and 2004.}
\label{fig:fading3}
\end{figure}

\subsubsection{The first part of the fading (until mid-1996)}

During the first year after peak brightness, OO\,Ser exhibited similar
exponential fading at all wavelengths. This can be quantified by
fitting a linear relationship to the first part of the light curves
plotted logarithmically in Fig.~\ref{fig:lightcurve2}, using data
taken only after peak brightness at a given wavelength. The derived
fading rates are displayed in Fig.~\ref{fig:fading3}~(a).
This graph shows that the fading rate during this period was
approximately $1.0\,{\pm}\,0.3$ mag/year irrespectively of wavelength
in the whole mid-infrared regime.

\subsubsection{The second part of the fading (until mid-1997)}
\label{sec:second}

The second part of the light curves (between 1996 October and 1997
September) can be characterised by fading rates of $0.35\,{\pm}\,0.10$
mag/year, similar at all wavelengths from $2.2$ to $60\,\mu$m
(Fig.~\ref{fig:fading3}~(b)). This is approximately three times slower
than the fading during the first part. In this period ISOPHOT-S
spectra were also obtained, which give the opportunity to analyse the
fading with high wavelength resolution. In order to increase the
signal-to-noise ratio, the fitted fading rates of ISOPHOT-S were
binned (in the case of the short wavelength channels) or smoothed (in
the case of the long-wavelength channels). The resulting values are
overplotted with small open or filled dots in
Fig.~\ref{fig:fading3}~(b), respectively. These values also support a
wavelength-independent fading, characteristic of the second part of
the outburst (except where the emission of OO\,Ser is strongly reduced
by the silicate absorption around $10\,\mu$m, introducing a large
uncertainty in the fit).

\subsubsection{The third part of the fading (until 2004)}
\label{sec:third}

New measurements from 2004 revealed that the fading monitored by
ISOPHOT continued at all wavelengths after 1997. We calculated fading
rates for the period 1997--2004. The resulting numbers, which are
plotted in Fig.~\ref{fig:fading3}~(c), are even lower than
those for the second part and they also show no
wavelength dependence. The values are in the range of
$0.19\,{\pm}\,0.08$ mag/year.

\subsubsection{The present status of OO\,Ser}
\label{sec:present}

In Fig.~\ref{fig:lightcurve2} the preoutburst fluxes (marked by dashed
lines) can be compared with the latest measurements from
2004-2006. This comparison reveals that in 2004 OO\,Ser was still
above the preoutburst state in the whole infrared wavelength regime,
indicating that the outburst had not finished yet. The same conclusion
can be drawn from Fig.~\ref{fig:sed} if one compares the SED from 1996
with that from 2004.

In Fig.~\ref{fig:lightcurve2} the last K-band photometric point
from 2006 seems to deviate from a linear extrapolation of the
preceding fading, possibly indicating a fourth phase of the outburst
with very slow flux change. Extrapolating the lightcurves using the
rates calculated for the third part of the fading (1997--2004),
the expected end date of the outburst is ${\approx}\,2011$ from both
the $2.2\,\mu$m and the $12\,\mu$m lightcurve. The slowing down
of the fading process, however, may delay this event even well beyond
2011.

\subsection{K-band spectral evolution}
\label{sec:spec}

Figure \ref{fig:spec} displays our K-band spectrum from 2006, together
with the outburst spectra from 1995 and 1998 \citep{hodapp99}. All
three spectra are taken with comparable slit widths and were
normalised to their value at $2.4\,\mu$m. In general, all spectra show
a steep continuum, rising towards longer wavelengths. No individual
absorption or emission lines can be seen, even in quiescence. The
overall shape of the three spectra is very similar, though there is a
tendency that later spectra are less steep.


\section{Discussion}
\label{sect:Discussion}
 
\subsection{The outburst}
\label{sec:out}

OO\,Ser offers a unique possibility to investigate the long-term
behaviour of an eruptive young stellar object. The rise time of
OO\,Ser was 8.5 months at the longest \citep{hodapp96}, similar to the
typical timescale of ``fast'' FUor outbursts \citep[e.g.~the outburst of
FU\,Ori or V1057\,Cyg, which was modelled by a triggered eruption,
see ][]{blhk}.

As discussed in Sec.~\ref{sec:2}, the fading rate abruptly decreased
350 days after the maximum brightness. A similar event was observed in
the case of V1057\,Cyg \citep{kh91} too, but there the transition
happened later, about 1300 days after the maximum. Using the light
curves of \citet{kh91} we computed fading rates for V1057\,Cyg and
compared them with corresponding values of OO\,Ser. We found that both
before and after the transition, the fading of OO\,Ser ($1.00$ mag/yr
before, $0.35$ mag/yr after) was significantly faster than that of
V1057\,Cyg ($0.09...0.40$ mag/yr before, $0.04...0.14$ mag/yr after,
between $0.44$ and $21\,\mu$m).

From the $2.2\,\mu$m light curve in Fig.~\ref{fig:lightcurve2} we can
conclude that the object will return to the quiescent phase some
time after 2011. This implies that the duration of the outburst of
OO\,Ser will be at least 16 years. This timescale differs both
from that of FUors (being several decades or a century) and that of
EXors (being some weeks or months), suggesting that OO\,Ser is a young
eruptive object that differs from both FUors and EXors.

The magnitude of the luminosity change during outburst is another
argument in favour of OO\,Ser being different from FUors or
EXors. From Fig.~\ref{fig:sed} we could estimate a bolometric
preoutburst luminosity of $L_{pre}\,{=}\,4.5\,{\pm}\,1.5 L_{\odot}$
for OO\,Ser and from the ISOPHOT data we also computed luminosities
for OO\,Ser for the different epochs during the outburst. Since the
far-infrared fluxes are contaminated by nearby sources, we calculated
an upper and a lower limit for the luminosity, by including or
neglecting the $100 - 200\,\mu$m data points, respectively. Adopting
the 1996 February values of $L_{peak}\,{=}\,26\,{\dots}\,36 L_{\odot}$
as a representative outburst luminosity range, OO\,Ser changed its
luminosity by a factor of about $6\,{\dots}\,8$. Thus one may conclude
that both the peak brightness and the amplitude of the luminosity
increase was significantly lower than the corresponding values of
classical FUors \citep[${\sim}\,100\,L_{\odot}$ and a factor of 100,
see][]{hk96}.

We note that there exists another star, V1647\,Ori, which seems to
share some characteristics of OO\,Ser. Based on the amplitude of
its brightening, V1647\,Ori was classified as an intermediate-type
object between FUors and EXors by \citet{muz}. Preoutburst and
outburst luminosities of V1647\,Ori are $L_{pre}\,{=}\,5.6\,L_{\odot}$
\citep{v1647ir} and $L_{peak}\,{=}\,44\,L_{\odot}$ \citep{muz},
respectively, thus its luminosity changed by a factor of about 8,
similarly to OO\,Ser. The timescales of their outbursts are somewhat
different, because the eruption of V1647\,Ori was only 2 years long
\citep{ibvs} and it also produced an outburst in the 1960s
\citep{aspin06}.

We speculate, following \citet{hodapp96}, that OO\,Ser (and in
some respects V1647\,Ori) may be the representative of a new class of
young eruptive stars \citep[``Deeply Embedded Outburst Star'' or
DEOS in][]{hodapp96}. Members of this class may be defined by their
relatively short timescales compared to FUors, possibly recurrent
outbursts, modest increase in bolometric luminosity and accretion
rate, and an evolutionary state earlier than that of typical FUors or
EXors (see Sec.~\ref{sec:evol}).

\subsection{The evolutionary stage of OO\,Ser}
\label{sec:evol}

Based on optical to millimetre measurements available at that time,
\citet{hodapp96} claimed that the SED of OO\,Ser appeared consistent
with that of a class I source and assumed its age to be $10^5\,$yr,
indicating that OO\,Ser is in a very early evolutionary phase. The new
data presented in this paper make it possible to reestablish the
evolutionary stage of the source. We followed the method of
\citet{chen} and computed the bolometric temperature
$T_{\mathrm{bol}}$ according to their Eq.~(1) for the preoutburst,
outburst and present SEDs. Due to the uncertainty of the far-infrared
data points (see Sec.~\ref{sec:conf} about source confusion), we
calculated $T_{\mathrm{bol}}$ in two different ways, with and without
data points affected by source confusion
($60\,{<}\,\lambda\,{<}\,800\,\mu$m). The resulting values are in the
range of 50 -- 120\,K for all three SEDs. Within this interval, the
bolometric temperature slightly increased, and the luminosity changed
by a factor of 7 during the outburst. We compared these values with
the distribution of corresponding values among young stellar objects
in the Taurus and $\rho$ Ophiuchus star forming regions
\citep{chen}. From this check we can conclude that OO\,Ser seems
to be an early class I object, and its age is ${<}\,10^5\,$yr.

\subsection{Structure of the circumstellar environment}

The circumstellar environment of young eruptive stars is usually
modelled with a flat or flared accretion disc and an extended
infalling envelope \citep[e.g.~][]{kh91, tbb}. In these models the
emission of the central source (the star and the innermost part of the
accretion disc) dominates the emission at optical and near-infrared
wavelengths. Between $3$ and $10\,\mu$m the origin of the emission is
the release of accretion energy in the disc, and also starlight
reprocessed in the same part of the disc and also in the envelope. The
emission at $\lambda\,{>}\,10\,\mu$m is starlight reprocessed in the
envelope. The outburst occurs when, due to thermal instability in the
inner edge of the disc, the accretion rate suddenly increases. After
peak brightness, the accretion slowly relaxes to its quiescent
value. The decreasing accretion rate causes the fading of the central
source and consequently leads to the simultaneous fading of the
reprocessing envelope. Thus, this model predicts a
wavelength-independent fading of the source in the whole optical to
mid-infrared regime (see e.g.~V1057\,Cyg in \citealt{fuors} and
V1647\,Ori in \citealt{jose}).

The circumstellar environment of OO\,Ser probably shares many
properties with the abovementioned models, except that the unusually
high extinction indicates a larger and/or denser envelope. As it can
be seen in Fig.~\ref{fig:fading3}, the fading of OO\,Ser was indeed
wavelength-independent in the whole near- to mid-infrared wavelength
regime, in agreement with the model predictions. This, together with
the overall shape of its SED and the ice features in its mid-infrared
spectrum indicates that the circumstellar structure of OO\,Ser is
similar to those of other young eruptive stars, i.e.~possesses a
circumstellar accretion disc and is embedded in a dense circumstellar
envelope.

\subsection{Viscosity}

\citet{bl94} modelled FUor outbursts as self-regulated accretion
events in protostellar accretion discs. In this model the risetime of
the outburst, the subsequent high state and the time between
successive outbursts are dependent on $\alpha$, the viscosity
parameter in the model of \citet{shakura}. Fitting their model to the
observed FUor timescales, \citet{bl94} derived
$\alpha_c\,{=}\,10^{-4}$ for the $\alpha$ value in the cool, neutral
state, and $\alpha_h\,{=}\,10^{-3}$ in the hot, ionised state.  Since
in many respects OO\,Ser is similar to FUors, the model of
\citet{bl94} might be applicable, although the timescales of the
OO\,Ser outburst are remarkably shorter than that of FUors. Thus,
applying this model to OO\,Ser requires different parameters than
those for FUors. \citet{bl94} give in their Table~2 the dependence of
different timescales on $\alpha$, from which we can estimate
$\alpha_h\,{=}\,10^{-2}$ for OO\,Ser.  This is one order of magnitude
higher than the usual value for FUors, which implies that OO\,Ser may
differ from classical FUors in a way that its disc has different, one
order of magnitude higher viscosity.


\section{Summary and conclusions}
\label{sect:Conclusions}

In this paper we presented an infrared monitoring programme on
OO\,Ser, a deeply embedded young eruptive star in the Serpens NW
star-forming region. OO\,Ser went into outburst in 1995 and has been
gradually fading since then. Our infrared photometric data obtained
between 1996 and 2006 revealed that the fading of the source is still
ongoing in the whole infrared wavelength regime, and that OO\,Ser will
probably not return to quiescent state before 2011. The flux
decay has become slower since the outburst peak and has been
practically wavelength-independent.

From these results we draw the following conclusions:
\begin{itemize}
\item The outburst timescale and the moderate luminosity suggest that
  OO\,Ser is different from both FUors and EXors, and shows some
  similarities to the recently erupted young star V1647\,Ori.
\item Based on its SED and bolometric temperature, OO\,Ser seems to be
  an early class I object, with an age of ${<}\,10^5\,$yr. As
  proposed by outburst models, the object is probably surrounded by an
  accretion disc and a dense envelope. This picture is also supported
  by the wavelength-independence of the fading.
\item Due to the shorter timescales, outburst models developed for
  FUors can only work for OO\,Ser if the viscosity parameter in the
  circumstellar disc, $\alpha$, is set to an order of magnitude higher
  value than usual for FUors.
\end{itemize}


\begin{acknowledgements} 
  The ISOPHOT data presented in this paper were reduced using the
  ISOPHOT Interactive Analisys package PIA, which is a joint
  development by the ESA Astrophysics Division and the ISOPHOT
  Consortium, lead by the Max-Planck-Institut f\"ur Astronomie
  (MPIA). We thank Gaspare Lo Curto for kindly providing us with the
  TIMMI2 data on OO\,Ser. We also thank the referee, Klaus Hodapp, for
  useful suggestions that greatly improved the paper. The work was
  partly supported by the grant OTKA K\,62304 of the Hungarian
  Scientific Research Fund. J.A.P. acknowledge support from grant AYA
  2001-1658, financed by the Spanish Direcci\'on General de
  Investigaci\'on.
\end{acknowledgements}


\bibliography{paper}
\bibliographystyle{aa}


\end{document}